Title:

Spiral Blurring Correction with Water-Fat Separation for Magnetic Resonance Fingerprinting in the Breast

Author names and degrees:

M.Sc. Teresa Nolte[1], M.Sc. Nicolas Gross-Weege[1], Dr. Mariya Doneva[2], Dipl. Ing. Peter Koken[2], M.Sc. Aaldert Elevelt[3], Dr. Daniel Truhn[4], Prof. Dr. Christiane Kuhl[4], Prof. Dr. Volkmar Schulz[1]

Author affiliations:

(1) Physics of Molecular Imaging Systems, Experimental Molecular Imaging, RWTH Aachen University, Germany
(2) Tomographic Imaging Systems, Philips Research Hamburg, Germany
(3) Oncology Solutions, Philips Research Eindhoven, The Netherlands
(4) Clinic for Diagnostic and Interventional Radiology, University Hospital Aachen, Germany

Corresponding author info:

Teresa Nolte, Center for Biohybrid Medical Systems, Forckenbeckstr. 55, 52074 Aachen, teresa.nolte@pmi.rwth-aachen.de, tel. +49 241 80 85911

- and-

Volkmar Schulz, Center for Biohybrid Medical Systems, Forckenbeckstr. 55, 52074 Aachen, volkmar.schulz@pmi.rwth-aachen.de, tel. +49 241 80 80116

Word count main (manuscript body): 4924 words

# Abstract/Keywords

PURPOSE: Magnetic Resonance Fingerprinting (MRF) with spiral readout enables rapid quantification of tissue relaxation times. However, it is prone to blurring due to off-resonance effects. Hence, fat blurring into adjacent regions might prevent identification of small tumors by their quantitative $T_1$ and $T_2$ values. This study aims to correct for the blurring artifacts, thereby enabling fast quantitative mapping in the female breast.

METHODS: The impact of fat blurring on spiral MRF results was first assessed by simulations. Then, MRF was combined with 3-point Dixon water-fat separation and spiral blurring correction based on conjugate phase reconstruction. The approach was assessed in phantom experiments and compared to Cartesian reference measurements, namely inversion recovery (IR), multi-echo spin echo (MESE) and Cartesian MRF, by normalized root mean square error (NRMSE) and standard deviation (STD) calculations. Feasibility is further demonstrated in-vivo for quantitative breast measurements of 6 healthy female volunteers, age range 24-31 years.

RESULTS: In the phantom experiment, the blurring correction reduced the NRMSE per phantom vial on average from 16% to 8% for T1 and from 18% to 11% for T2 when comparing spiral MRF to IR/MESE sequences. When comparing to Cartesian MRF, the NRMSE reduced from 15% to 8% for T1 and from 12% to 7% for T2. Furthermore, STDs decreased. In-vivo, the blurring correction removed fat bias on T1/T2 from a rim of about 7-8 mm width adjacent to fatty structures.

CONCLUSION: The blurring correction for spiral MRF yields improved quantitative maps in the presence of water and fat.

# Introduction

Quantitative Magnetic Resonance Imaging (qMRI) offers an vendor independent imaging contrast, which promises the identification and classification of lesions based on their intrinsic tissue properties[1,2]. Moreover, quantitative image data represents an optimal input for post processing, such as machine learning methods[3]. The tissue relaxation times $T_1$ and $T_2$ are intrinsic tissue parameters that underlie the contrast formation of the clinically used qualitative, i.e., contrast weighted, MR images. However, the acquisition of quantitative parameter maps has not yet widely found its way into clinical practice, mainly due to long scan times.

In breast imaging, previous reports suggest that qMRI can help to determine the response to neoadjuvant chemotherapy[4-6] (namely, decreased T2 values are reported for responders) as well as to distinguish invasive ductal carcinoma from healthy tissue[7] or between different types of lesions[8,9]. Moreover, if the observed differences prove to be significant, a fast quantitative breast imaging protocol may be of interest in contrast-agent free breast screening.

Magnetic Resonance Fingerprinting (MRF) is a fast sequence that measures several quantitative markers at a time[10,11] from an image series with varying acquisition parameters such as flip angles, repetition times and RF phases. The measured signal evolution in every voxel is compared to a dictionary of simulated signal evolutions, which permits to select the best-matching quantitative parameters. MRF allows for highly efficient parameter estimation, as the MRF signal is acquired during transient state while making use of high undersampling during readout in each TR interval. Up to a certain undersampling factors / for long enough MRF sequences, correct identification of the underlying tissue properties is possible as long as the resulting undersampling artifacts distribute in a noise-like manner around the true signal evolution[11-13]. Spiral readout is often preferred for MRF because of its sampling speed and large k-space coverage[14].

However, spiral sampling results in blurred images for off-resonant spins. This effect becomes especially important if the field of view (FOV) does not only contain aqueous tissues, but also fat, of which the main spectral peak presents an average chemical shift of about -3.5 ppm with respect to the resonance frequency of water[15]. In consequence, fat signal that has blurred into adjacent voxels obscure the contours of tissues as well as the presence of adjacent structures of interest, such as small tumors. A conjugate phase reconstruction (CPR) allows correction for off-resonance induced blurring artefacts in spiral images[16]. Yet, CPR requires knowledge about the spatial off-resonance distribution. A different approach that circumvents fat blurring is suppression of the fat signal, e.g. by fat saturation techniques[17]. Fat-saturated MRF was recently presented in the abdomen as well as in female breast[7,18]. However, fat saturation techniques may not always yield complete suppression of the fat signal over the entire FOV, especially at higher field strengths and/or in breast MRI protocols that involve larger FOVs to cover both breasts such as the axial bilateral imaging protocols used for breast cancer screening. In the female breast anatomy, the fat signal may provide diagnostic information as well. For instance, keeping the fat signal in $T_2$-weighted images (and rather not suppressing it) permitted the distinction of benign from malignant tumors in such lesions that showed enhancement during dynamic contrast enhanced MRI[19].

In this work, we extend MRF by a Dixon water-fat separation approach[20], which allows to correct for the fat blurring. The presented method does not require the separate acquisition of an off-resonance map. It is inspired by the approach of Boernert et al.[21], who combined a 3-point Dixon method with CPR on fully sampled spiral images. CPR can equally deblur undersampled MRF data [22,23]. In both cases, the authors characterized the off-resonance map in a separate scan before computing the CPR of the individual, undersampled images. Preliminary results on fat blurring-corrected MRF with water-fat separation were recently presented[24-25]. Very recently, alternative MRF methods estimating water $T_1$ and $T_2$/water $T_1$ and fat $T_1$ as well as the fat signal fraction were proposed[26,27]. However, the breast anatomy has not yet been addressed.

We thus propose 2D blurring-corrected MRF with Dixon water-fat separation in the female breast, where both aqueous fibroglandular tissue as well as fatty tissue are present. Thereby, quantitative parameter maps of the relaxation times in the breast as well as the off-resonance map are obtained.

# Methods

## CPR for spiral off-resonance blurring correction

Spiral MRI is prone to off-resonance artifacts. Deviations $\Delta\omega$ from the water proton resonance frequency $\omega_0$ may result from the spatial inhomogeneity of the main magnetic field, i.e., due to local differences in magnetic susceptibility, or from the chemical shift of a tissue, as in the case of fat. If a spin distribution $\rho(\boldsymbol{r})$ is subject to any type of spatially varying off-resonance frequency $\Delta\omega(\boldsymbol{r})$, the MR signal can be written as

$$S(t) = \int \rho(\boldsymbol{r}) \cdot \exp(-i\Delta\omega(\boldsymbol{r})t) \exp\big(-i\boldsymbol{r}\boldsymbol{k}(t)\big) \, d\boldsymbol{r}. \qquad [1]$$

For reconstruction, the spiral signal $S(t)$ is commonly interpolated onto a Cartesian k-space grid prior to Fourier transformation into the image space[28], i.e., both $\boldsymbol{r}$ and $\boldsymbol{k}$ are defined on a Cartesian grid. According to Eq. 1, the true spin distribution $\rho(\boldsymbol{r})$ accrues an extra phase term during signal readout if $\Delta\omega(\boldsymbol{r}) \neq 0$. Thus, standard image reconstruction by inverse Fourier transform results in a blurred image $M(\boldsymbol{r}) = \mathcal{F}^{-1}\big(S(\boldsymbol{k})\big)$, as $\rho(\boldsymbol{r})$ is convolved by the spiral point spread function in the image space[29]. However, if the off-resonance map $\Delta\omega(\boldsymbol{r})$ as well as the spiral trajectory $\boldsymbol{k}(t)$ are known, the blurring-free $\rho(\boldsymbol{r})$ can be well approximated[16] by calculating the CPR

$$\rho^{CPR}(\boldsymbol{r}) = \frac{1}{(2\pi)^2} \int S(\boldsymbol{k}) \exp\big(i\Delta\omega(\boldsymbol{r})\tau(\boldsymbol{k})\big) \exp(i\boldsymbol{r}\boldsymbol{k}) \, d\boldsymbol{k}. \qquad [2]$$

Here, $\tau(\boldsymbol{k})$ is the inverted spiral k-space trajectory $\boldsymbol{k}(t)$, i.e., a map that indicates the time at which a k-space location $\boldsymbol{k} = (k_x, k_y)$ is reached. Numerical implementation of the CPR comprises the following steps: (1) compute $\tau(\boldsymbol{k})$ from the gradient shapes, (2) transform the blurred image $M(\boldsymbol{r})$ into k-space, (3) demodulate each pixel of location $\boldsymbol{r} = (x, y)$ at the corresponding off-resonance frequency $\Delta\omega(\boldsymbol{r})$. Step (3) can be accelerated by demodulating $S(\boldsymbol{k})$ with an array of discrete, evenly spaced off-resonance frequencies[30]. Here, an array of equidistant frequencies $\Delta\omega_i = 2\pi \cdot [-200, 200]$ rad is used to compute 201 demodulations of the blurry image. For each location $\boldsymbol{r}$, the deblurred pixel value is chosen from the image with the demodulation frequency $\Delta\omega_i$ that is closest to $\Delta\omega(\boldsymbol{r})$. Acceleration is important, knowing that MRF requires computing the CPR for a large series of images.

If both aqueous and fatty tissues are present in the FOV, they experience different off-resonance frequencies: A distribution of water protons $\rho^w(\boldsymbol{r})$ sees only the the inhomogeneities of the main magnetic field: $\Delta\omega^w(\boldsymbol{r}) = \Delta\omega_0(\boldsymbol{r})$. In contrast, for a distribution of fat protons $\rho^f(\boldsymbol{r})$, the off-resonance frequency is shifted by -3.5 ppm: $\Delta\omega^f(\boldsymbol{r}) = \Delta\omega_0(\boldsymbol{r}) + \Delta\omega_{cs}$. At 1.5 T, the chemical shift of fat with respect to water is $\Delta\omega_{cs} = -2\pi \cdot 220$ Hz.

## MRF-Dixon with spiral deblurring

To correct for off-resonance blurring in MRF, we combined spiral MRF with a 3-point Dixon water-fat separation and CPR deblurring. For superposing signal fractions of water and fat in the same voxel, the resulting voxel signal $M(\boldsymbol{r}) = M^w(\boldsymbol{r}) + M^f(\boldsymbol{r})$, acquired at TE, may be written as

$$M(\boldsymbol{r}) = \left[\rho^w(\boldsymbol{r}) + \rho^f(\boldsymbol{r}) \cdot \exp(i\,\Delta\omega_{cs}TE)\right] \cdot \exp(i\Delta\omega_0(\boldsymbol{r})TE) \cdot \exp\left(i\phi_0(\boldsymbol{r})\right). \qquad [3]$$

Here, $\phi_0$ is a constant receiver offset-phase. For simplicity, we use a single peak fat model, although fat exhibits multiple spectral components. In three-point Dixon methods, three complex images $M_q(\boldsymbol{r})$ ($q = \{1,2,3\}$) of different echo times $TE_q$ are acquired and serve to recover $\rho^w(\boldsymbol{r})$, $\rho^f(\boldsymbol{r})$ and $\Delta\omega_0(\boldsymbol{r})$. In our MRF-Dixon approach, we select $TE_q = (2\pi/\omega_0) + (q-1) \cdot \Delta TE$, with $\Delta TE = \pi/\omega_0$, which allows for an analytical solution of the water-fat separation[15].

The MRF-Dixon acquisition and post-processing are sketched in Figure 1. The MRF sequence is a gradient-spoiled (i.e., unbalanced) gradient echo sequence. Three MRF trains of $N$ pulses are played out, which are separated by a delay time $\Delta t_d$. Spiral acquisition begins after each RF excitation at $TE_q$ for the q-th MRF train. In result, $3 \cdot N$ undersampled complex images are acquired, with water and fat signal in-phase, out-of-phase and in-phase again for the first, second and third MRF train, respectively.

First, the off-resonance map is retrieved from the undersampled MRF data. Temporal averages over each of the three MRF trains are calculated, which highly reduces the undersampling induced aliasing artifacts that are present in the individual images:

$$\bar{M}_q = \sum_{j=1}^{N} M_{qj} \qquad [4]$$

The mean off-resonance map

$$\overline{\Delta\omega_0} = 2 \cdot \Delta TE \cdot \arg\left(\frac{\overline{M_3}}{\overline{M_1}}\right) \qquad [5]$$

is calculated and phase unwrapping is applied to $\overline{\Delta\omega_0}$ if phase jumps are present within the breast. Phase unwrapping was implemented as a region-growing algorithm[15]. The unwrapped off-resonance map is then used to execute a 3-point Dixon water-fat separation on every individual time point $j=1...N$ of the MRF acquisition, following Eq. 5. Hence, a blurred water-only and a fat-only MRF train are retrieved. As the next post-processing step, a CPR with $\Delta\omega = \overline{\Delta\omega_0}$ is conducted on every complex image of the water-only MRF train, whereas a CPR with $\Delta\omega = \overline{\Delta\omega_0} + \Delta\omega_{cs}$ is conducted on every complex image of the fat-only MRF train. After CPR calculation, the deblurred water-only and fat-only dataset are recombined, i.e., added up, and subsequently matched to an MRF dictionary of simulated signal evolutions.

## Experimental

### Simulation study: MRF with off-resonance blurring

We conducted a simulation study to estimate the impact of spiral off-resonance blurring on the MRF relaxation times of structures near fatty tissue in the breast. The simulation phantom (size 224 × 224 voxels, 430 cm square FOV) contained a ring of fatty tissue ($\Delta\omega = \Delta\omega_{cs}$) and an adjacent small test structure (TS) ($\Delta\omega = 0$, square of 5 × 5 voxels, 1 voxel/1.92 mm distance from the fat border), both surrounded by fibroglandular (FG) tissue ($\Delta\omega = 0$). The nominal T1 and T2 maps in Figure 2(a) and (d) show the 100 × 100 voxel region containing the structures of interest. We simulated MRF signal evolutions of the three tissues as well as a full MRF dictionary based on the extended phase graph (EPG) formalism[31] with MATLAB (The MathWorks, Natick, United States). EPG simulations employed the FA sequence[32] depicted in Figure 1(a) preceded by an 180° inversion pulse, an unbalanced gradient in slice selection and repetition/echo times of TR/TE=(20/4.6) ms. The $T_1$ and $T_2$ resolution of the dictionary was as stated in Table 1. Hence, the fat signal was deliberately blurred using equation [2] with $\Delta\omega = -\Delta\omega_{cs}$ and spiral k-space trajectories of different acquisition time $T_{acq}$. After matching the blurred simulation data to the MRF dictionary, we examined line profiles of $T_1$ and $T_2$ through the TS as well as the mean relaxation time values within the TS and their standard deviations.

### Phantom validation

To validate the MRF-Dixon acquisition, a water-fat phantom, i.e., 8 vials with mixtures of gelatin and varying amounts of a Gd-based contrast agent embedded in lard (pig fat), was prepared and scanned next to a 1 L bottle of CuSO4/water solution. Phantom and in-vivo breast MR scans were acquired on a 1.5 T system (Achieva, Philips, Best, The Netherlands) with a 4-channel breast coil (Invivo Corporation,

Gainesville, Florida) in axial orientation. The acquired scans and their durations are stated in Table 2. For MRF-Dixon scans, a square FOV of 430 mm size with voxels of $(1.92 \times 1.92 \times 5)$ mm$^3$ was selected. As in the simulation study, we utilized a constant TR of 20 ms and the train of 500 flip angles[32] depicted in Figure 1(a), preceded by a 180° inversion pulse. Echo times (TE$_1$/TE$_2$/TE$_3$) = (4.61/6.92/9.23) ms were set for the three MRF trains, corresponding to in-phase/out-of-phase/in-phase readout at 1.5 T. The delay time in between the MRF trains was set to $t_d$ = 7.5 s to allow for complete relaxation of the magnetization in breast tissues. A single spiral interleaf of uniform sampling density (acquisition window $T_{acq}$=7ms) was acquired in each TR interval, corresponding to an undersampling factor of R = 20. Between successive TR intervals, the k-space trajectory was rotated by 18 degrees. The transmit field (B$_1$$^+$) inhomogeneity over the slice was measured in a separate Cartesian 3D sequence using the actual flip angle technique[33]. The MRF-Dixon dataset was deblurred based on the above-described approach. To compare between different sampling strategies, Cartesian MRF data was further acquired with TE=4.61 ms. To retrieve T$_1$ and T$_2$ parameter maps, a dictionary with approximately 300.000 normalized entries was calculated. B$_1$$^+$ inhomogeneity was included in the dictionary as a multiplicative correction factor f$_{B1+}$ in front of the flip angle train. The dictionary resolution is specified in Table 1. To reconstruct T$_1$ and T$_2$ maps, the measured signal evolution in every voxel was first normalized to a complex magnitude of 1 and then compared to the subset of dictionary entries with f$_{B1+}$ closest to the measured B$_1$$^+$ of that voxel. The best matching dictionary entry was selected based on the maximum inner product between dictionary entry and measured signal evolution[10]. To evaluate the effect of CPR deblurring on the matching results, matching was equally performed to the first MRF-train $M_1$ without any correction for blurring, equal to the standard MRF measurement and matching procedure[10].

All Cartesian reference scans were acquired with a reduced FOV of 80% in right-left direction to shorten the overall scan time. An additional SENSE factor of 1.5 was intrinsically applied in the scanner reconstruction to avoid fold-over artifacts, e.g. from the arms in the breast scans. The readout bandwidth was maximized for the Cartesian scans, corresponding to an actual fat shift of 0.127 px. Quantitative Cartesian reference measurements, i.e., inversion recovery (IR) for $T_1$ and multi-echo spin echo (MESE) for $T_2$, were acquired and compared to the results of the MRF matching. IR measurements in the phantom employed 11 inversion times TI=(50, 100, 200, 400, 600, 800, 1100, 1500, 2000, 3000, 5000), a turbo factor of 16 and TR/TE = (10000/3.5) ms. $T_1$ values were retrieved for every voxel by fitting the function

$$M(TI) = M_0 \cdot \left| 1 - 2 \cdot \exp\left(-\frac{TI}{T_1}\right) + \exp\left(-\frac{TR}{T_1}\right) \right| \qquad [6]$$



to the time series of IR images. For the MESE sequence in the phantom, $n = 1 \ldots 30$ images with $TE_n = n \cdot 35$ ms and TR = 10000 ms were acquired. $T_2$ values were retrieved for every voxel by fitting the function

$$M(TE) = M_0 \cdot \exp\left(-\frac{TE}{T_2}\right) \qquad [7]$$

to the time series of MESE images. Circular regions of interest (ROI) covering the phantom vials were defined and the standard deviations of T1 and T2 within each phantom vial were calculated. For each phantom vial, Normalized Root Mean Square Errors (NRMSE) were calculated between the spiral MRF and the reference sequences:

$$\text{NRMSE}(A, B) = \sqrt{\sum_{j \in ROI} \frac{T_{ij}^A - T_{ij}^B}{T_{ij}^B}} \qquad [8]$$

Here, $i = \{1, 2\}$. "A" stands for either the Standard MRF or the MRF-Dixon measurement, while "B" stands for either the IR/MESE or the Cartesian MRF measurement.

### In-vivo breast scans

Breast MR scans were acquired of six female healthy volunteers after informed consent, with age and ACR breast density as stated in Table 3. The breasts were immobilized in cranio-caudal direction.

As in the phantom, an undersampled spiral MRF-Dixon sequence was acquired (R=20). In order to verify the robustness of our MRF-Dixon acquisition in-vivo to undersampling artifacts and hence the quality of the parameter maps, a fully sampled MRF measurement (R=1) was performed for 3 out of the 6 volunteers. A Cartesian 3-point Dixon sequence was performed as a reference for water-fat separation and to validate our CPR deblurring correction. Echo times ($TE_1/TE_2/TE_3$) = (1.42/2.92/4.42) ms were used, thereby maximizing signal-to-noise ratio (SNR) with a phase accrual of 120 degrees between successive echoes. TR was set to 1000 ms and the flip angle was set to 20°. The scanner software reconstructed images of the water and the fat signal as well as an off-resonance map, based on an iterative least squares approach[34] and a multi-peak spectral model of fat.

Reference IR measurements in the volunteers employed 12 inversion times TI = (100, 200, 300, 400, 500, 600, 800, 1000, 1300, 1600, 2000, 2300) ms, a turbo factor of 10 and TR/TE = (3000/4.61) ms. For the $T_2$ reference measurement, $n = 1 \ldots 30$ images with $TE_n = n \cdot 9.22$ ms and TR = 3000 ms were acquired.

# Results

## Simulation study

Figure 2 shows the results of the off-resonance blurring simulations. In subfigures 2(a)-(f), MRF $T_1$ and $T_2$ maps including fat blurring are exemplarily depicted for spiral $T_{acq}$ of 7 ms and 16 ms, next to the nominal maps. For both $T_{acq}$, blurred fat signal smears out of the fatty tissue region and affects the relaxation times within the test structure (TS). Line profiles through the TS are shown in subfigures 2(g) and (h). Already for short $T_{acq}$, fat shifts over the contour of the TS, thereby dissimulating it. For the later used $T_{acq}$=7 ms, blurred fat signal spreads over about 4 pixels, i.e., about 7 to 8 mm. With increasing $T_{acq}$, the quantitative $T_1$ and $T_2$ values in the TS become increasingly biased towards the fat relaxation times. Moreover, the fat blurring adds variability to the values. These two effects become equally visible in subfigures 2(i) and (j), which depicts mean value and standard deviation of $T_1$ and $T_2$ within the TS for different spiral $T_{acq}$.

## Phantom validation

Figure 3 shows the validation of the MRF-Dixon sequence in the phantom. Subfigures 3(a) and (b) depict the parameter maps resulting from IR/MESE reference measurements, Cartesian MRF, standard (i.e., blurred) spiral MRF and the proposed MRF-Dixon approach. While the phantom vials in the standard MRF maps exhibit fat blurring artefacts, these are greatly reduced in the MRF-Dixon maps. In subfigure 3(b), it can be seen that both spiral and the Cartesian MRF measurements underestimate the T2 MESE values for large T2 values. Subfigures 3(c) and (f) depict the standard deviation (STD) over the phantom vials for all four measurements. In the MRF-Dixon measurement, all STDs are reduced with respect to Standard MRF for both T1 and T2, although Cartesian MRF and the IR/MESE reference sequences show even smaller STDs. Subfigures 3(d)-(h) depict the NRMSEs calculated according to Eq. 8, calculated per vial for T1 and T2. Not only with respect to the IR/MESE reference sequences, but also with respect to the Cartesian MRF sequence, the MRF-Dixon sequence yields smaller NRMSE values than the Standard MRF sequence. This holds true for both T1 and T2. Average NRMSE values over the eight phantom vials for T1 are (16, 8, 15, 8)% between (Standard MRF and IR reference, MRF-Dixon and IR reference, Standard MRF and Cartesian MRF, MRF-Dixon and Cartesian MRF). For T2, corresponding values of (18, 11, 12, 7)% are calculated.

## In-vivo breast scans

As an example for the breast scans, we present the full dataset for one volunteer. Further results are available in the Supporting Information Figures S1 and S2.

### Deblurring and water-fat separation

Figure 4 presents two maps of the background off-resonance $\Delta\omega_0$ of the main magnetic field. Subfigure 4(a) shows the mean off-resonance map ($f_0$-map) that we calculated from the MRF-Dixon measurement by using Eq. 5 and a subsequent phase unwrapping step. Subfigure 4(b) shows the $f_0$-map as obtained from the Cartesian Dixon reference measurement. The mean $f_0$-map computed from the MRF-Dixon signals is free from phase wraps and does not show any artifacts from spiral sampling. In one spot that is marked by a white arrow, the two maps differ: the Cartesian Dixon reference map shows a local maximum, while the MRF-Dixon map does not. This discrepancy is also visible in the difference map in subfigure 4(c), which else exhibits values mostly between 0 and -15 Hz. This figure also reveals a more structured appearance of the MRF-Dixon map with respect to the (smoothed) Cartesian map.

Mean MRF signals, i.e., the temporal averages over the water-fat separated MRF trains, were calculated according to Eq. 4. Figures 5(a) and (d) show the mean water and fat signal, respectively, prior to CPR deblurring. While the mean water signal shows little blurring, the mean fat signal is strongly smeared out. This makes the anatomical features hardly distinguishable. Subfigures 5(b) and (e) show the mean water and fat signal after CPR deblurring. Deblurring alters the mean fat signal most strongly, resulting in a fat distribution with sharp edges that permits to delineate the same anatomical features as seen in the anatomical reference scan, cf. Figure 2. The changes in the mean water signal due to deblurring are more subtle, but the deblurred mean water signal in subfigure 5(b) reveals, for example, a sharper delineation of the skin. When comparing the deblurred images to the Cartesian Dixon water and fat image shown in subfigures 5(c) and (f), respectively, a close resemblance is observed: the same features are visible with a similar degree of sharpness. Yet, there is one visible difference between both measurements, namely in the same location that differed already in the $f_0$-maps. The local maximum in the Cartesian $f_0$-map results in a smaller fat signal and higher water signal in that location, which we indicated again by white arrows. The discrepancy is likely due to the smoothness constraint used in the computation of the Cartesian Dixon map and needs further investigation.

### Quantitative parameter maps

Figure 6 presents the results for the quantitative parameter maps. Subfigures 6(a) and (b) show the standard MRF matching to the first MRF train $M_1$, without any correction for spiral blurring. It is clearly visible that the fat blurring propagates into the $T_1$ and $T_2$ map. Specifically, areas of fatty tissue appear broadened and without any clear delineation to the adjacent fibroglandular tissue. This broadening also results in a breast size that is extending over the anatomical breast size. Furthermore, streak artifacts of circular shape are present in the $T_2$ maps. The $T_1$ and $T_2$ maps after deblurring are shown in subfigures 6(c) and (d) for the undersampled MRF-Dixon measurement and in subfigures 6(e) and (f) for the fully sampled MRF measurement, respectively. After CPR deblurring, the MRF matching yields improved parameter maps showing a sharp delineation between fibroglandular and fatty tissue. Fatty substructures within the breast are now clearly visible and the contour of the outer fat layer is no longer smeared out. When comparing the undersampled MRF-Dixon measurement to the fully sampled MRF-Dixon measurement, the $T_1$ and $T_2$ maps look very similar, indicating stability of our MRF-Dixon sequence to undersampling. However, the undersampled $T_2$ maps are generally more noisy than their fully sampled counterparts. In the $T_2$ map in subfigure (d), a slightly streaky structure is visible also after deblurring. The $T_2$ map reconstructed from the fully sampled MRF-Dixon measurement is completely free of the aforementioned artifacts, as can be seen in subfigure (f). As the MRF matching was corrected for the measured $B_1^+$, the reconstructed $T_2$ maps are free of asymmetry, i.e., they show similar $T_2$ values for the left and right breast. Subfigures 6(g) and (h) present the results of the $T_1$ and $T_2$ relaxometry measurements that we acquired for reference. Subfigure 6(g) shows the $T_1$ map as obtained from fitting Eq. 6 to the IR measurements. Subfigure 6(h) shows the $T_2$ map as obtained from fitting Eq. 7 to the MESE measurement. Both reference maps exhibit the same overall features as the parameter maps obtained by the MRF-Dixon method. However, the $T_2$ reference values of fatty tissue exhibit a positive offset with respect to the MRF-Dixon measurements. In addition, the $T_1$ reference values in fibroglandular tissue are overall smaller than the MRF-Dixon values.

To further assess the impact of fat blurring on small feature relaxometry, absolute difference maps between standard MRF measurement and MRF-Dixon measurement are shown in Figure 7(a) and (c) for $T_1$ and $T_2$, respectively. In both difference maps, a rim of altered $T_1$ and $T_2$ values is visible next to the fat border within the fibroglandular tissue. Within this zone, extending over about 4 pixels or about 7 to 8 mm, bias is added to the relaxation times and small features may be obscured. The line profiles depicted in subfigures 7(b) and (d) show changes in FG tissue of about 300 ms for $T_1$ and of about 30 ms for $T_2$. As well, the fat blurring outside the breast yielding a larger apparent breast size is visible from the line profiles.

# Discussion and Conclusions

This work addresses the blurring problem in spiral MRF for water and fat by a three-point Dixon approach. Three fingerprint trains of different echo time permit both water-fat separation and deblurring without requiring a separate off-resonance map. Thereby, an accurate measurement of the relaxation times of small features by spiral MRF becomes possible in regions that are else compromised by the overlapping, blurred fat signal.

In the simulation study, we first investigated the resulting bias on $T_1$ and $T_2$ near a fatty structure, depending on the spiral acquisition time. In the phantom validation experiments, we observed smaller NRMSE values between MRF-Dixon and Cartesian reference relaxation times than between Standard MRF and the reference. We equally see this improvement when calculating the NRMSEs with respect to Cartesian MRF. It should be underlined that the latter comparison judges the effect of our blurring correction best, as the spiral and Cartesian MRF sequences were employing equal acquisition parameters apart from the signal readout. The difference in long T2 values between MRF and MESE measurements is likely attributable to increased diffusion effects in the MRF sequence[35]. However, we do not expect such large $T_2$ in breast tissues[36]. We therefore conclude that the validation of the MRF sequence was relevant with respect to the intended application.

For the in-vivo breast scans, the deblurring approach via CPR resulted in blurring-free mean water and fat signals. The effect of deblurring was most prominent for the fat signal, as the scanner's resonance frequency usually adjusts close to the water resonance frequency. Retrospectively, the successful deblurring justifies using the mean off-resonance map during CPR, despite minor differences to the Cartesian Dixon map. Deblurring further permitted an improved feature delineation in both the $T_1$ and the $T_2$ maps. The quantitative maps of the undersampled MRF-Dixon measurement agreed well with those of the fully sampled one, despite the 20-fold acceleration. Next to the phantom measurements, this is an important indicator for the stability of our MRF-Dixon sequence in the presence of undersampling. We suggest that such a comparison should be made each time that an MRF sequence is changed, especially if the amount of acquired information is decreased, e.g. when reducing the number of TR intervals or the voxel sizes.

Future effort will comprise removal of the streak artifacts, which are supposedly due to wrong registration of signal in the presence of heart movement and through-plane blood flow. While for Cartesian sampling the in-flowing blood results in coherent ghosts along the phase-encoding direction[37] the spiral readout, bearing a continuously changing phase-encoding direction, smears such signal around the source of flow in a spiral-looking manner. A solution to this problem might lie in presaturation of signal in the heart region.

A different strategy may be to increase the signal-to-noise ratio (and thus to decrease the importance of flow artifacts) during reconstruction, such as by compressed sensing[38] or matrix completion methods[39].

We corrected the presented MRF-Dixon measurements for in plane $B_1^+$ inhomogeneity. Slice profile effects were not corrected during MRF matching; however, we employed an RF pulse shape with a time-bandwidth product of 10.2 that minimizes slice profile effects. $B_1^+$ correction proved to remove the large intra-breast inhomogeneity of the $T_2$ values[40]. MRF is known to be prone to $B_1^+$ inhomogeneity[41], as the dictionary reconstruction relies on exact knowledge of the flip angle train. Admittedly, a faster $B_1^+$ mapping method would be preferred for future MRF exams.

In-vivo, differences were present between the relaxation times in the MRF and the reference maps. The MRF and reference pulse sequences differed in the employed gradients and RF pulse shapes, which complicates their direct comparison. Slice profile effects and imperfect inversion pulses[42-43], diffusion[35,45] and magnetization transfer effects[46,47] are confounding factors affecting both MRF and reference relaxation measurements to different degrees, which can explain the differences in the relaxation time maps. In addition, fat has multiple spectral components with different relaxation times. This may lead to different apparent relaxation times for different sequences. These discrepancies are a problem yet to be solved by qMRI, which we cannot remedy by our deblurring approach alone.

Three separate MRF trains of different echo time are demonstrated here as a proof of principle that the approach works. Although we were still able to acquire a single slice in less than a minute, prolonged scan times will be of concern for volumetric acquisitions which are of relevance for breast imaging. Acceleration can be achieved by performing only a two-point Dixon water-fat separation with an additional phase-unwrapping step[48]. Instead of acquiring two or three separate MRF trains, several spirals may be acquired in one TR interval. For multiple slices, the delay times between MRF trains can moreover be used to acquire another slice. It should be mentioned that MRF in the breast with fat suppression as proposed by Chen et al.[7] is advantageous with respect to scan time, as only one MRF train is needed. A different strategy for water-fat separation without the need for several echo trains can lie in dictionary-based methods[27], which may afterwards be combined with spiral deblurring.

Deblurring of MRF data was demonstrated for six young healthy volunteers, presenting different breast densities. Blurring was removed both in case of demarcated fibroglandular/fat interfaces and for more distributed mixtures of fibroglandular and fatty tissue. Due to the technical feasibility nature of the study, the measurements do not reflect overall demography and ACR distribution in women. However, our maps already suggests that there may be a high variability in breast $T_1$ values – partially caused by partial volume effects where fat and water are present in the same voxel, but also especially within areas of fibroglandular tissue only, where most breast carcinoma can be found. This challenges tissue quantification in the breast.

Finally, we would like to point out that the separated water- and fat MRF data may be used to compute the fat signal fraction per voxel, either from the water and fat proton densities or by utilizing the mean water and fat signals. As this would add one more diagnostic parameter to the outcome of MRF, we aim to compare and validate these approaches in a future study.

Tables

**Table 1:** Parameter ranges and resolution of the MRF dictionary.

| Parameter | Range | Step size |
|---|---|---|
| $T_1$ / ms | [5,200] | 5 |
| | [210,500] | 10 |
| | [520,2000] | 20 |
| $T_2$ / ms | [2,100] | 2 |
| | [105,200] | 5 |
| | [210,500] | 10 |
| $f_{B1}$ | [0.7, 1.3] | 0.025 |

**Table 2:** Scan durations for phantom and in-vivo scans. Scans marked with "-" were not acquired. US = undersampled, FS = fully sampled, R = acceleration factor with respect to full sampling, IR = inversion recovery, MESE = multi-echo spin echo.

| Scan | Scan duration, phantom experiment | Scan duration, in-vivo experiments |
|---|---|---|
| Spiral MRF-Dixon, US, R=20 | 53 s | 53 s |
| Spiral MRF-Dixon, R=10 | 1 min 45 s | - |
| Spiral MRF-Dixon, R=5 | 3 min 30 s | - |
| Spiral MRF-Dixon, FS, R=1 | 17 min 31 s | 17 min 31 s |
| Cartesian MRF | 52 min 34 | - |
| B1 map (3D) | 3 min 38 s | 3 min 38 s |
| IR measurements | 22 min 0 s | 12 min 0 s |
| MESE measurements | 30 min 20 s | 9 min 6 s |
| Cartesian 3-point Dixon | 3 min 17 s | 3 min 17 s |

**Table 3:** Age and ACR breast density of the volunteers.

| Volunteer | Age | ACR breast density |
|-----------|-----|--------------------|
| 1 | 25 | 2 |
| 2 | 27 | 3 |
| 3 | 24 | 4 |
| 4 | 26 | 4 |
| 5 | 31 | 3 |
| 6 | 28 | 3 |

Figures

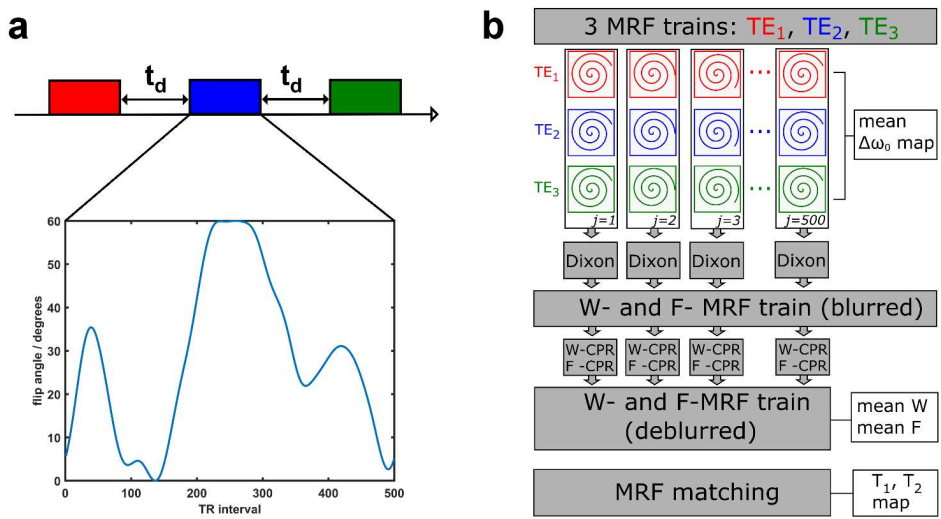

**Figure 1.** (a) MRF-Dixon acquisition. Three MRF trains of j=1...500 TR intervals each are acquired, differing in their echo time TE. The flip angle trains are separated by a delay time $t_d$. The flip angle train with a length of 500 TR intervals is shown in the bottom left. It is preceded by a 180° inversion pulse and has previously been published by Sommer et al.[32] (b) MRF-Dixon post-processing scheme. As a first step, the mean off-resonance map is computed from the temporal averages of the in-phase MRF trains ($TE_1$, $TE_3$). With the mean off-resonance map, a 3-point Dixon water-fat separation is conducted for each TR interval. A blurred water MRF train and a blurred fat MRF train are obtained, which are subsequently deblurred by CPR. By calculating the temporal average over the deblurred water and fat MRF train, we obtain a mean water and a mean fat image. In a last step, the deblurred water and fat data are recombined and subjected to the MRF matching process. In result, deblurred $T_1$ and $T_2$ maps are obtained.

**MRF T1 maps**

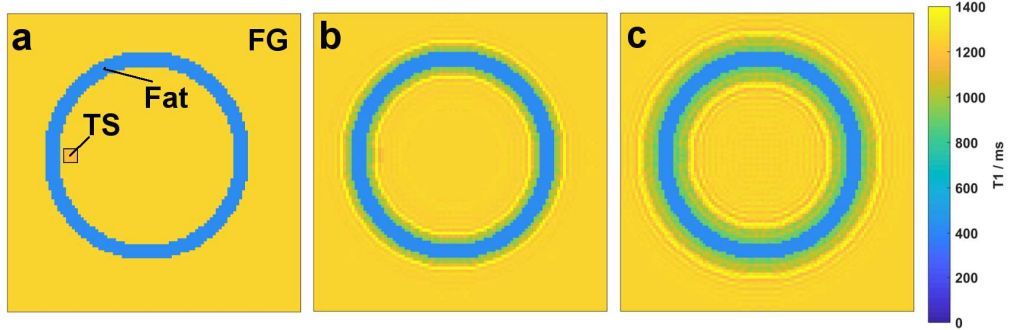

**MRF T2 maps**

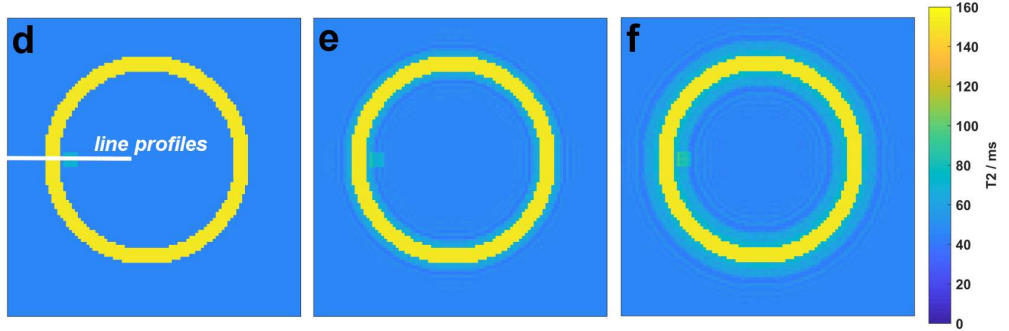

**Line profiles**

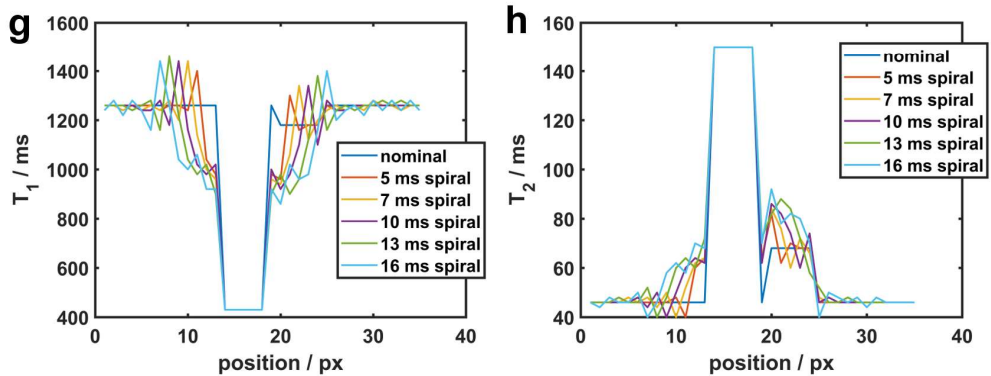

**Mean difference (TS to FG)**

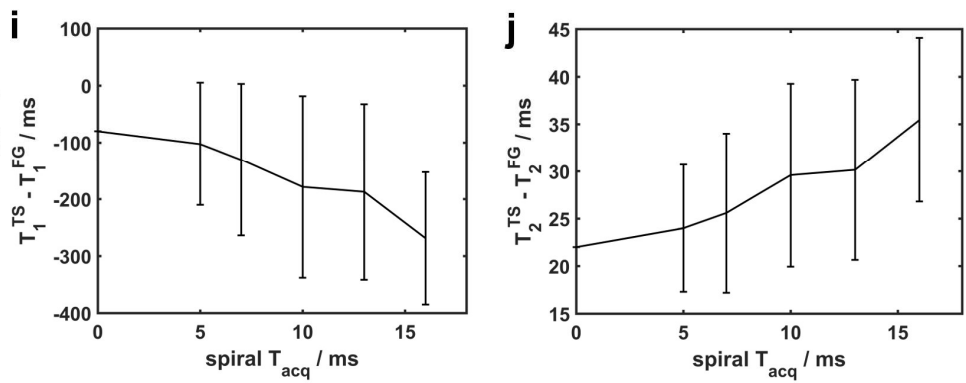

**Figure 2**: Simulation study for a phantom containing fibroglandular tissue (FG), a small test structure (TS) of slightly different relaxation times and a ring-shaped fatty structure. (a) and (b) show the nominal MRF $T_1$ and $T_2$ maps without any spiral fat blurring, (b) and (e) show the MRF maps after

simulation of fat blurring for a spiral of acquisition time $T_{acq}$=7 ms, (c) and (f) show the MRF maps after simulation of fat blurring for a spiral of $T_{acq}$ = 16 ms. Line profiles through the test structure for different spiral $T_{acq}$ between 5 ms and 16 ms are shown in (g) for $T_1$ and (h) for $T_2$. (i) and (j) show the difference between $T_1$ and $T_2$ mean values within the test structure and the background fibroglandular tissue as well as the standard deviations of the $T_1$ and $T_2$ values over the test structure for the different spiral $T_{acq}$.

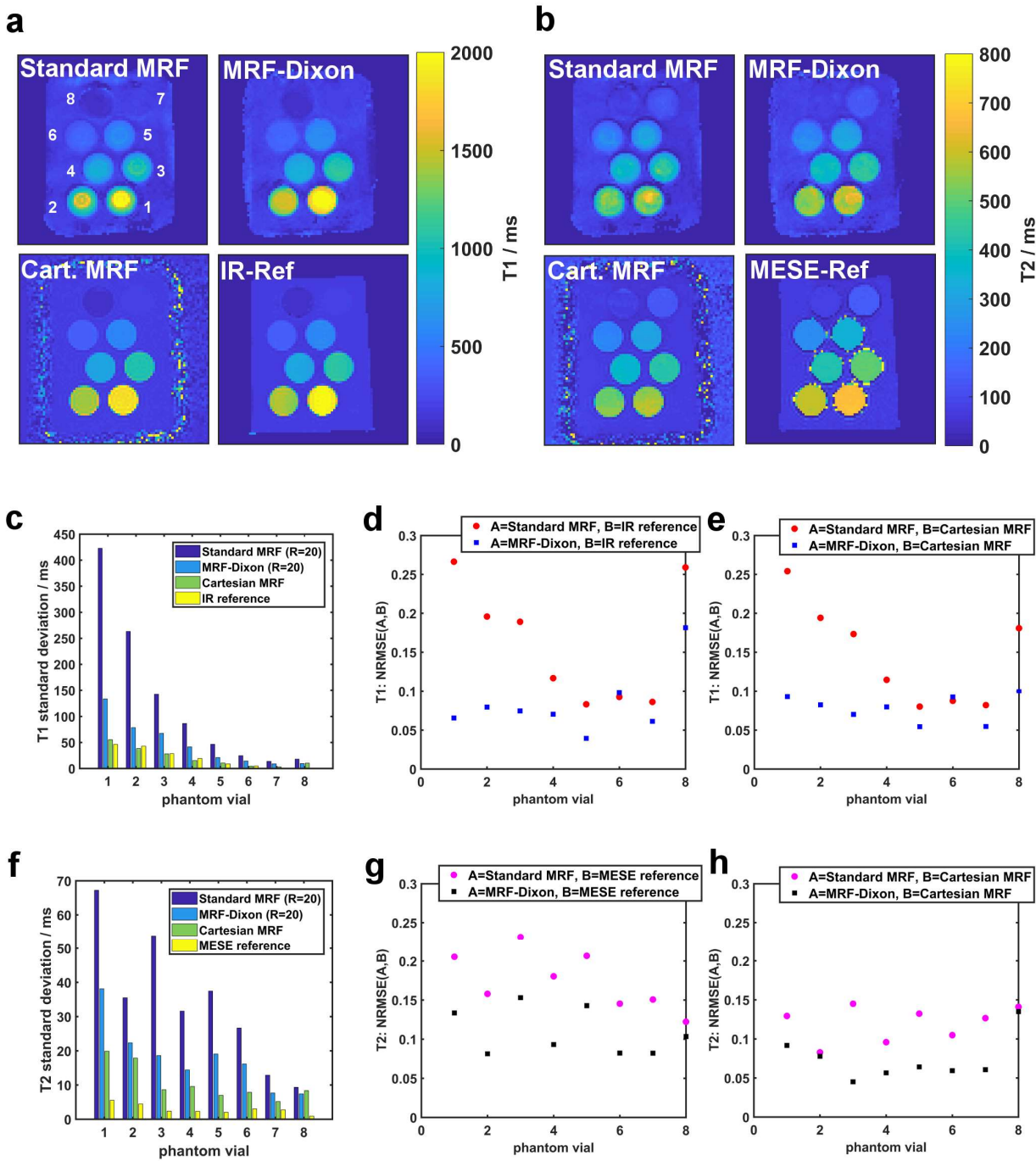

**Figure 3**: Validation of the MRF-Dixon sequence in a phantom. The phantom consists of eight vials of gelatin mixed with different amounts of Gadolinium embedded in lard (pig fat). Subfigures (a) and (b) show T1 and T2 maps of the phantom measurements. Depicted are 100 x 100 voxel large zooms onto the phantom. Both spiral MRF measurements, i.e., the (uncorrected) standard MRF measurement and the (blurring-corrected) MRF-Dixon measurement, employed an undersampling factor of R=20.

Subfigure (c) shows the standard deviation of the T1 values in each phantom vial for the Standard MRF measurement, the MRF-Dixon measurement, the Cartesian MRF measurement and the IR reference measurement. Subfigure (d) shows the T1-NRMSE between Standard MRF/MRF-Dixon and IR reference measurement. Subfigure (e) shows the T1-NRMSE between Standard MRF/MRF-Dixon and the Cartesian MRF measurement. Subfigure (f) shows the standard deviation of the T2 values in each phantom vial. Subfigure (g) shows the T2-NRMSE between Standard MRF/MRF-Dixon and IR reference measurement. Subfigure (h) shows the T2-NRMSE between Standard MRF/MRF-Dixon and the Cartesian MRF measurement.

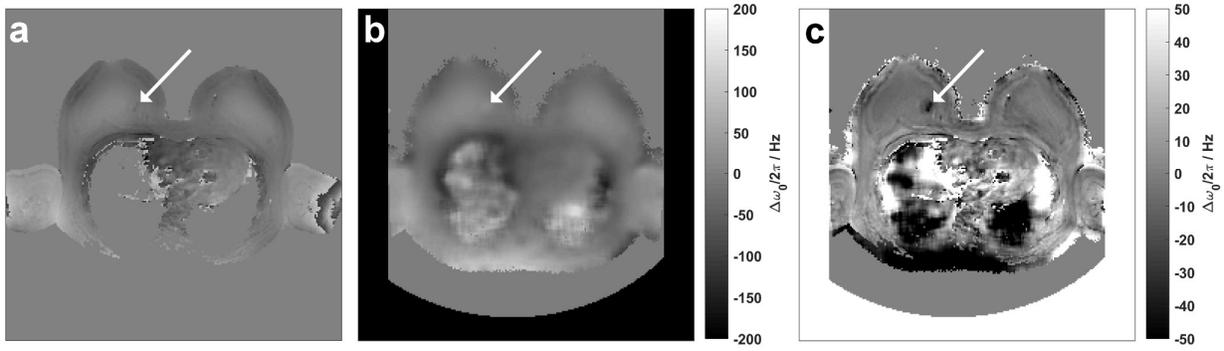

**Figure 4.** Off-resonance maps. (a) Mean off-resonance map as computed from the MRF-Dixon measurement according to Eq. 5. (b) Off-resonance map as reconstructed by the scanner's 3-point Dixon sequence. The white arrow points out a location where (a) and (b) differ. (c) Difference map, i.e., (a) – (b).

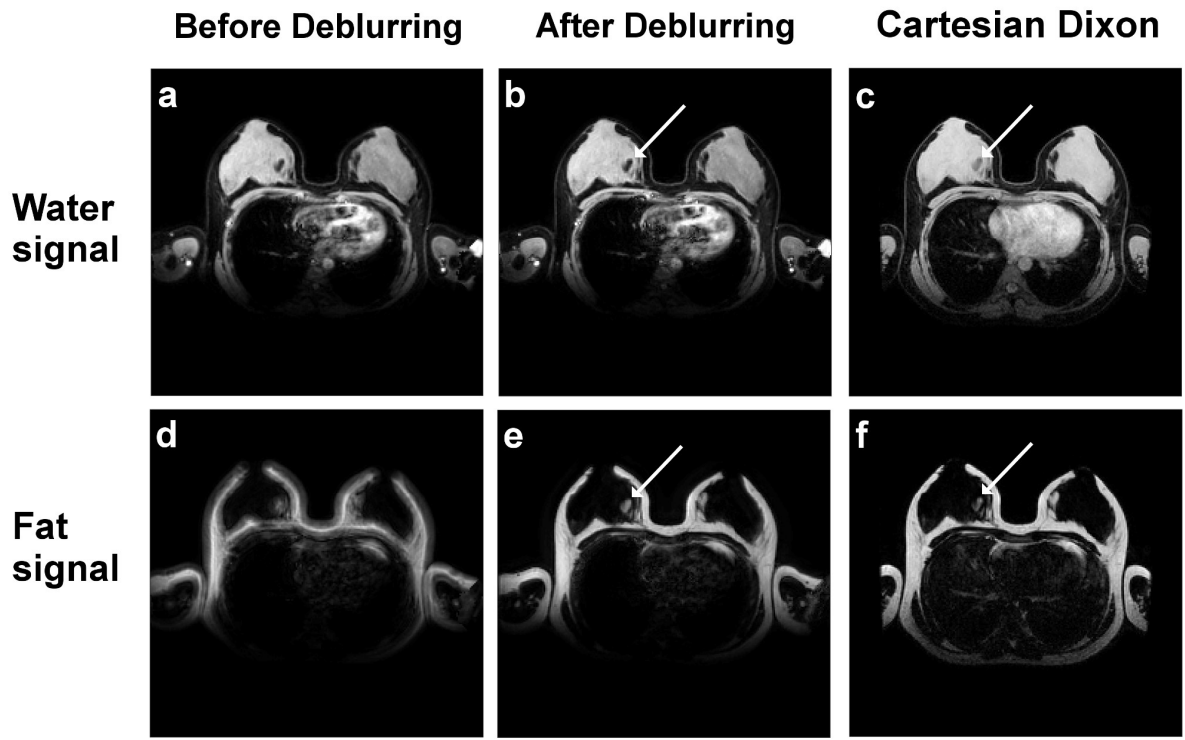

**Figure 5**: Deblurring results and comparison to the reference scan. (a) Mean MRF water signal and (d) mean MRF fat signal after Dixon water-fat separation, prior to CPR deblurring. (b) Mean MRF water signal and (e) mean MRF fat signal after Dixon water-fat separation, after CPR deblurring. (c) Cartesian 3-point Dixon water and (f) fat signal as obtained from the scanner reconstruction software.

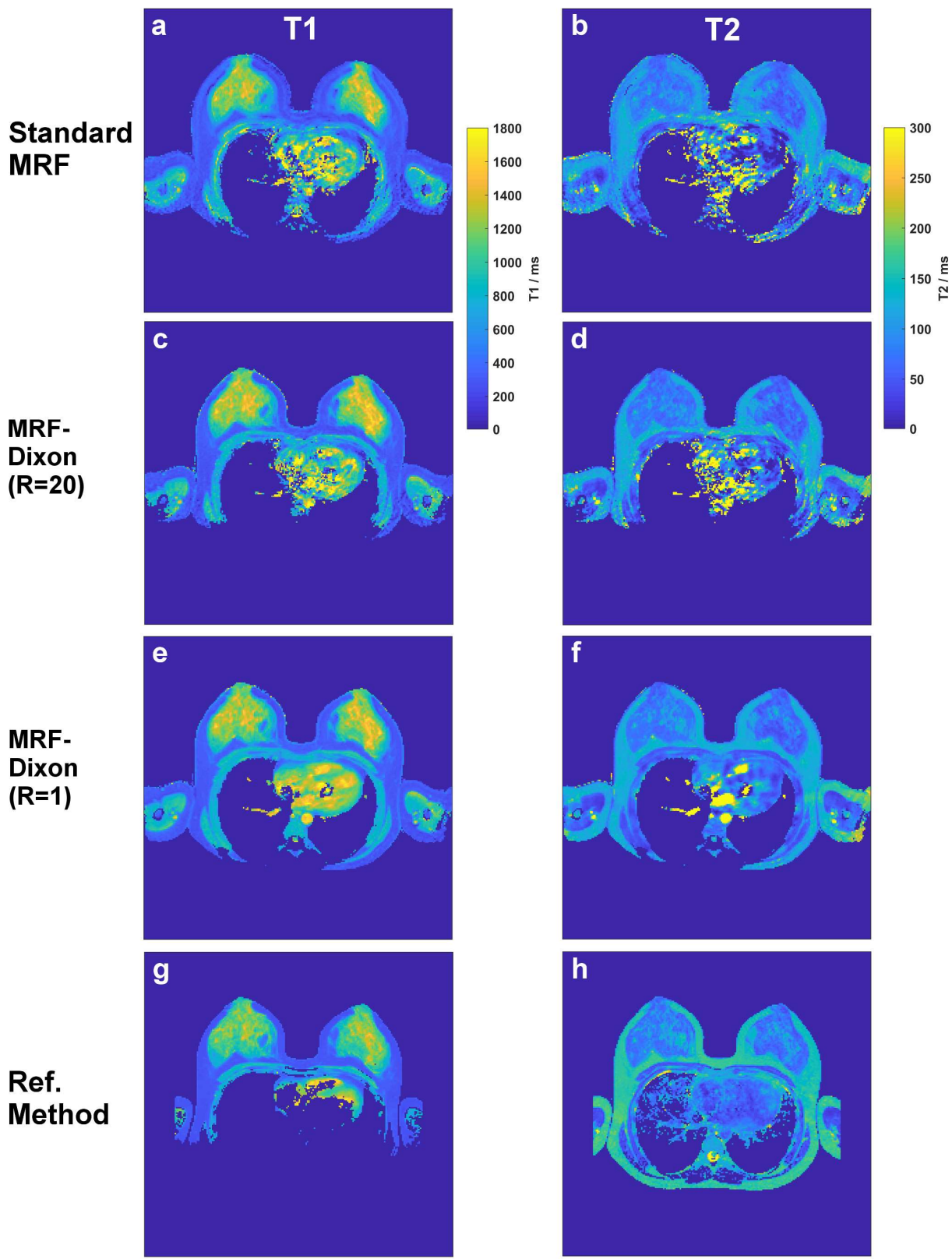

**Figure 6**: Relaxation time maps. (a) and (b) show the blurry $T_1$ and $T_2$ map as obtained from the standard MRF approach, i.e., when matching only the first out of the three MRF trains to the dictionary. (c) and (d) show the deblurred $T_1$ and $T_2$ map as obtained from the undersampled (R=20) MRF-Dixon measurement. (e) and (f) show the deblurred $T_1$ and $T_2$ map as obtained from the fully sampled (R=1) MRF-Dixon measurement. (g) and (h) show the $T_1$ and $T_2$ map as obtained from the reference methods, i.e., the inversion recovery and multi-echo spin echo measurement.

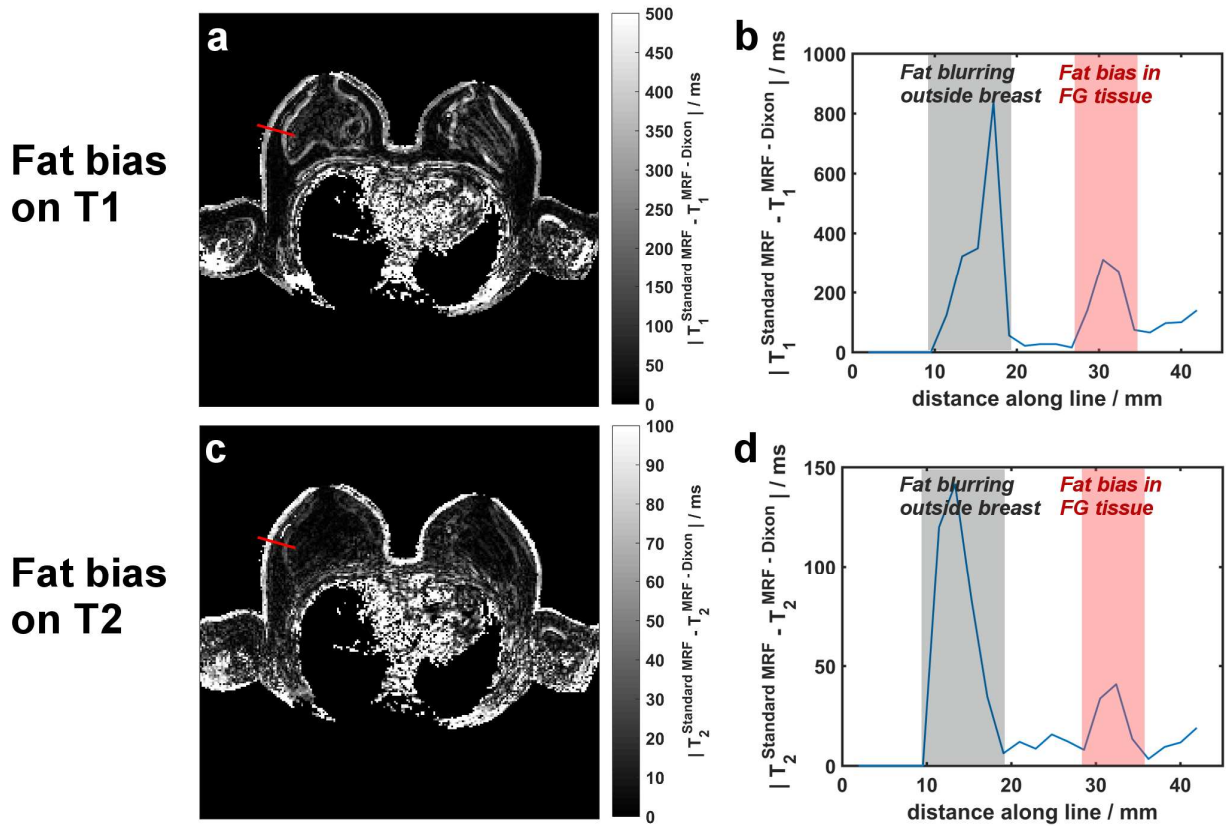

**Figure 7**: (a) Absolute T1 difference map between the (blurry) Standard MRF measurement and the (deblurred) MRF-Dixon measurement. The fat blurring manifests as a rim around the fatty structures. (b) Absolute T1 difference profile along the red line marked in (a). (c) Absolute T2 difference map between the Standard MRF measurement and the MRF-Dixon measurement. (d) Absolute T1 difference profile along the red line marked in (c). It is visible from the profiles that fat blurring causes bias within the fibroglandular tissue along a distance of about 4 pixels, i.e., about 7 – 8 mm.

**Supporting material figure S1:** Deblurring and water-fat separation results for volunteers 1, 2, 3, 5 and 6. The first column shows the mean water and fat signal as obtained from the MRF-Dixon measurements without deblurring. The second column shows the mean water and fat signal as obtained from the MRF-Dixon after deblurring. The third column shows the water and fat signal as obtained by the Cartesian Dixon reference measurement.

(a) Volunteer 1

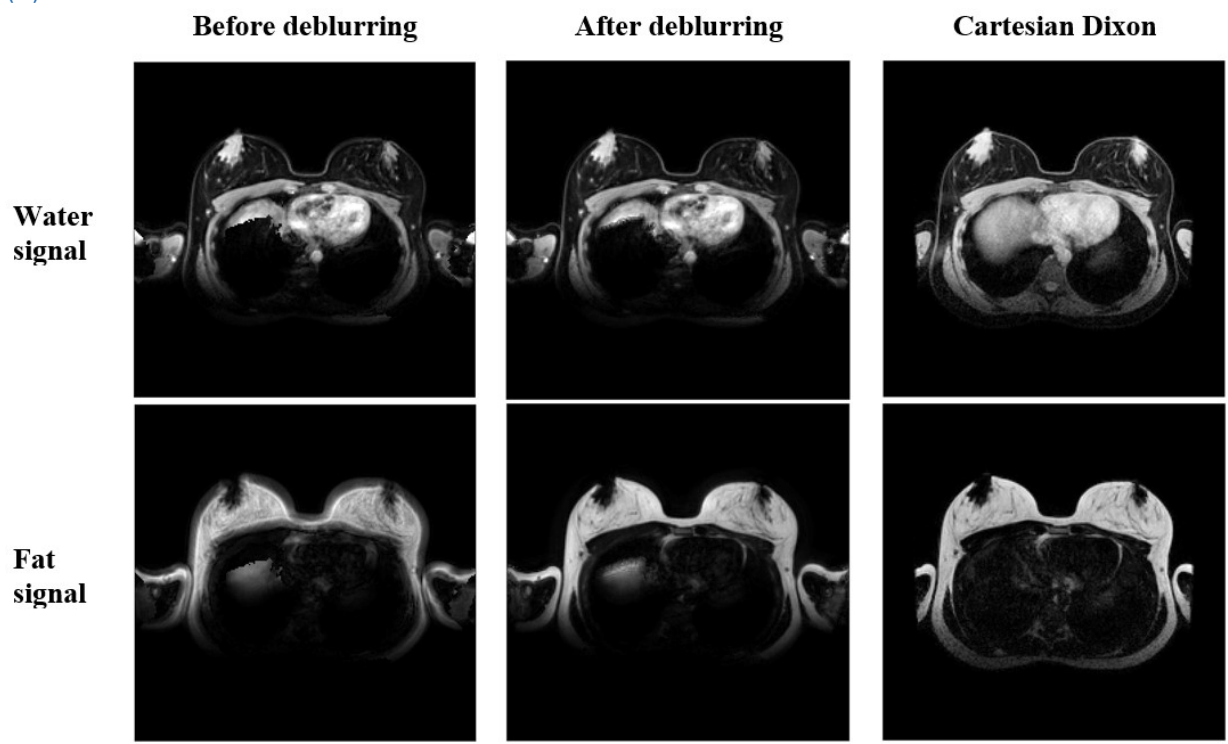

(b) Volunteer 2

|   | **Before deblurring** | **After deblurring** | **Cartesian Dixon** |

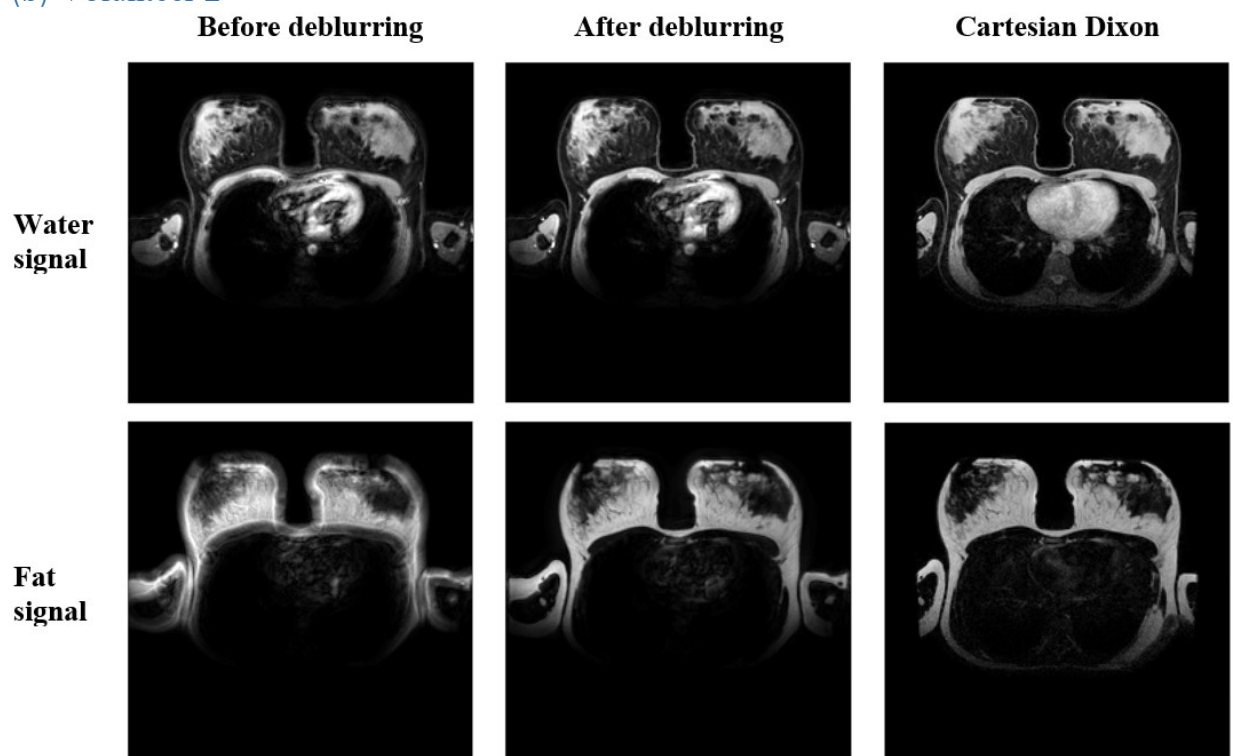

**Water signal**

**Fat signal**

(c) Volunteer 3

|   | **Before deblurring** | **After deblurring** | **Cartesian Dixon** |

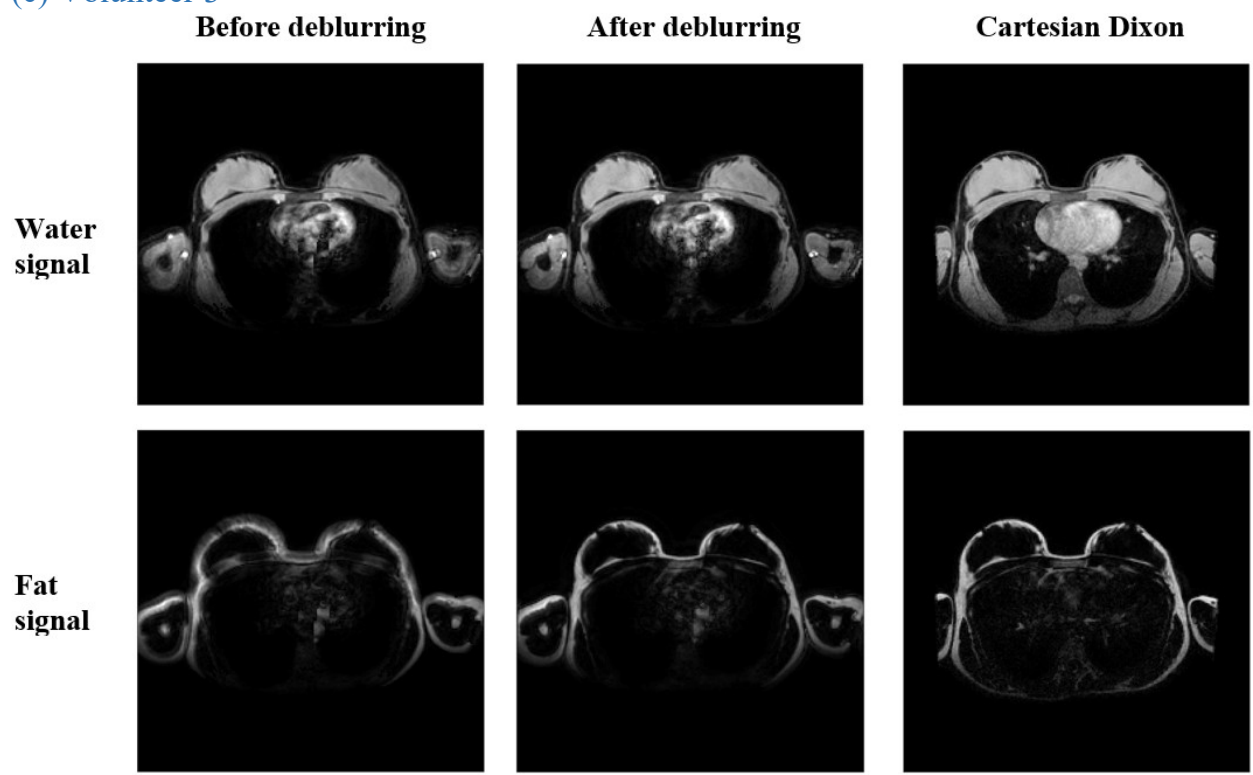

**Water signal**

**Fat signal**

(d) Volunteer 5

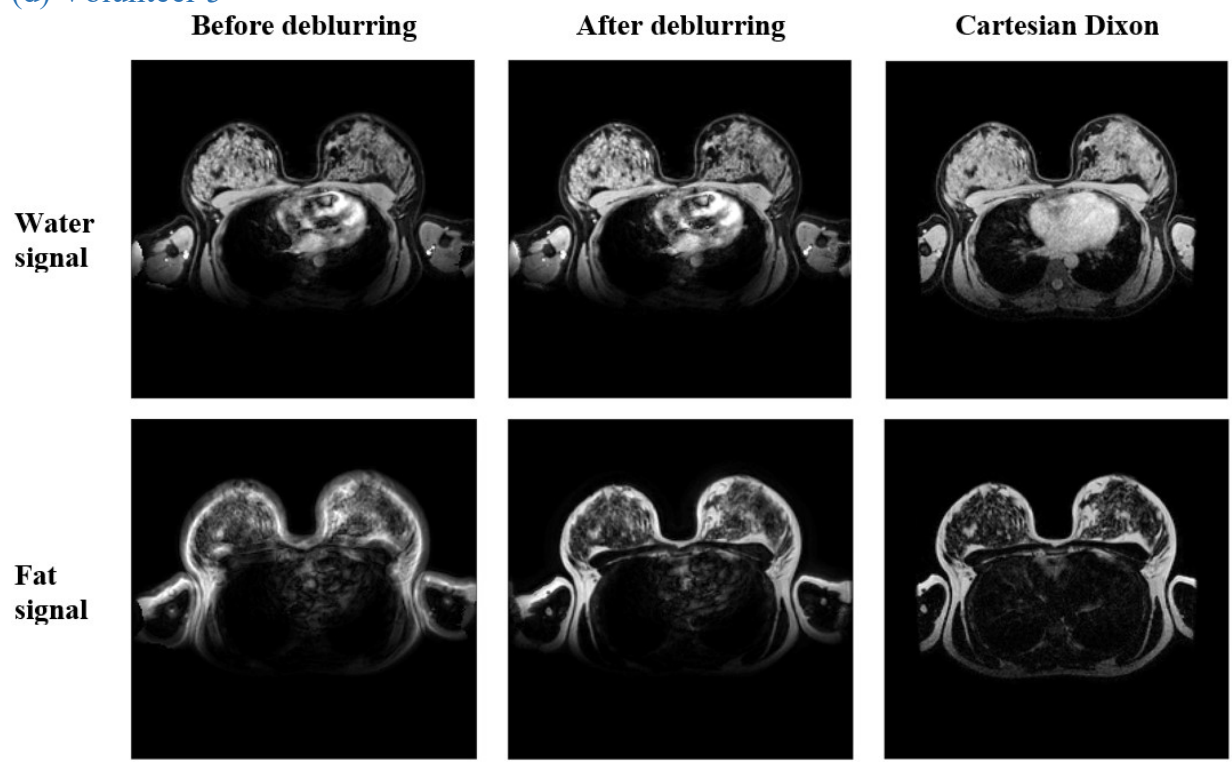

(e) Volunteer 6

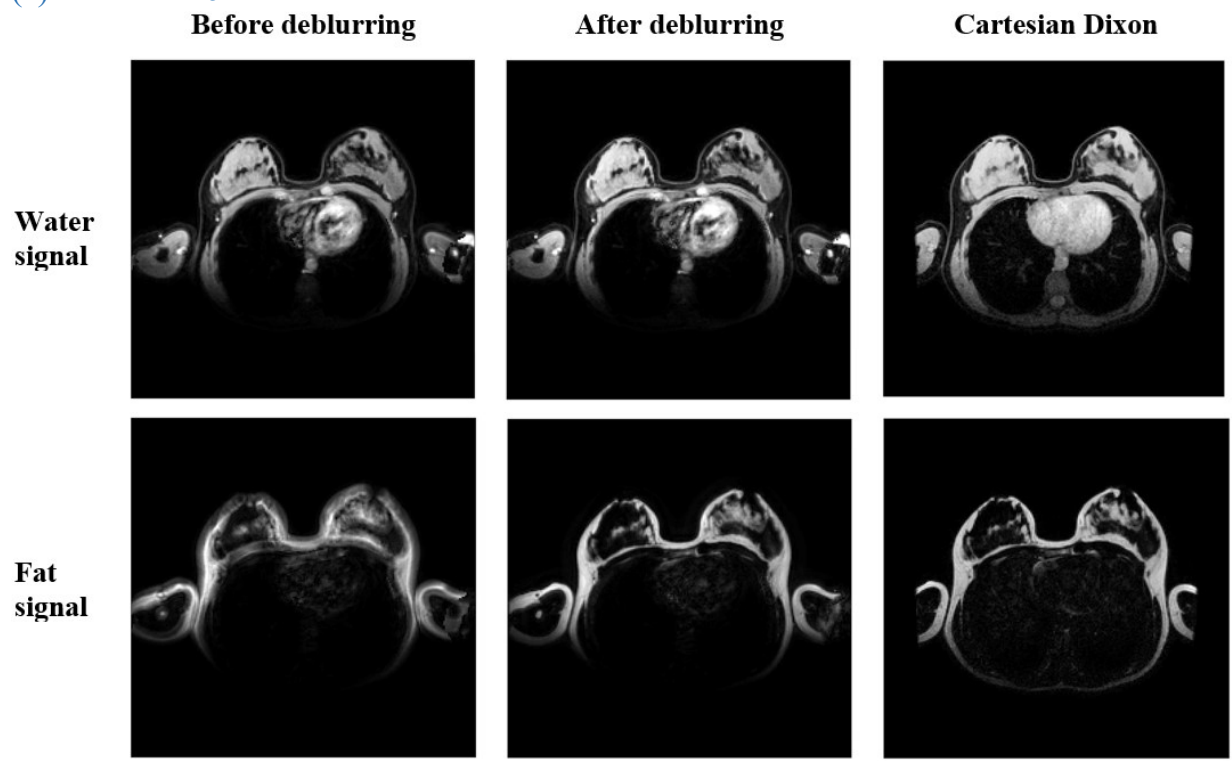

**Supporting material figure S2:** Relaxation time maps for volunteers 1,2,3,5 and 6. Top row: $T_1$ and $T_2$ map as obtained from the standard MRF matching to one single MRF train, i.e. without deblurring. Second (and third) row: $T_1$ and $T_2$ map as obtained from the undersampled (and the fully sampled, if acquired) MRF-Dixon measurement after deblurring. The fully sampled measurement was not acquired for volunteers 1, 3 and 6 due to the long total scan time. Bottom row: $T_1$ and $T_2$ map as obtained from the reference methods, i.e., inversion recovery for $T_1$ and multi-echo spin echo for $T_2$.

(a) Volunteer 1

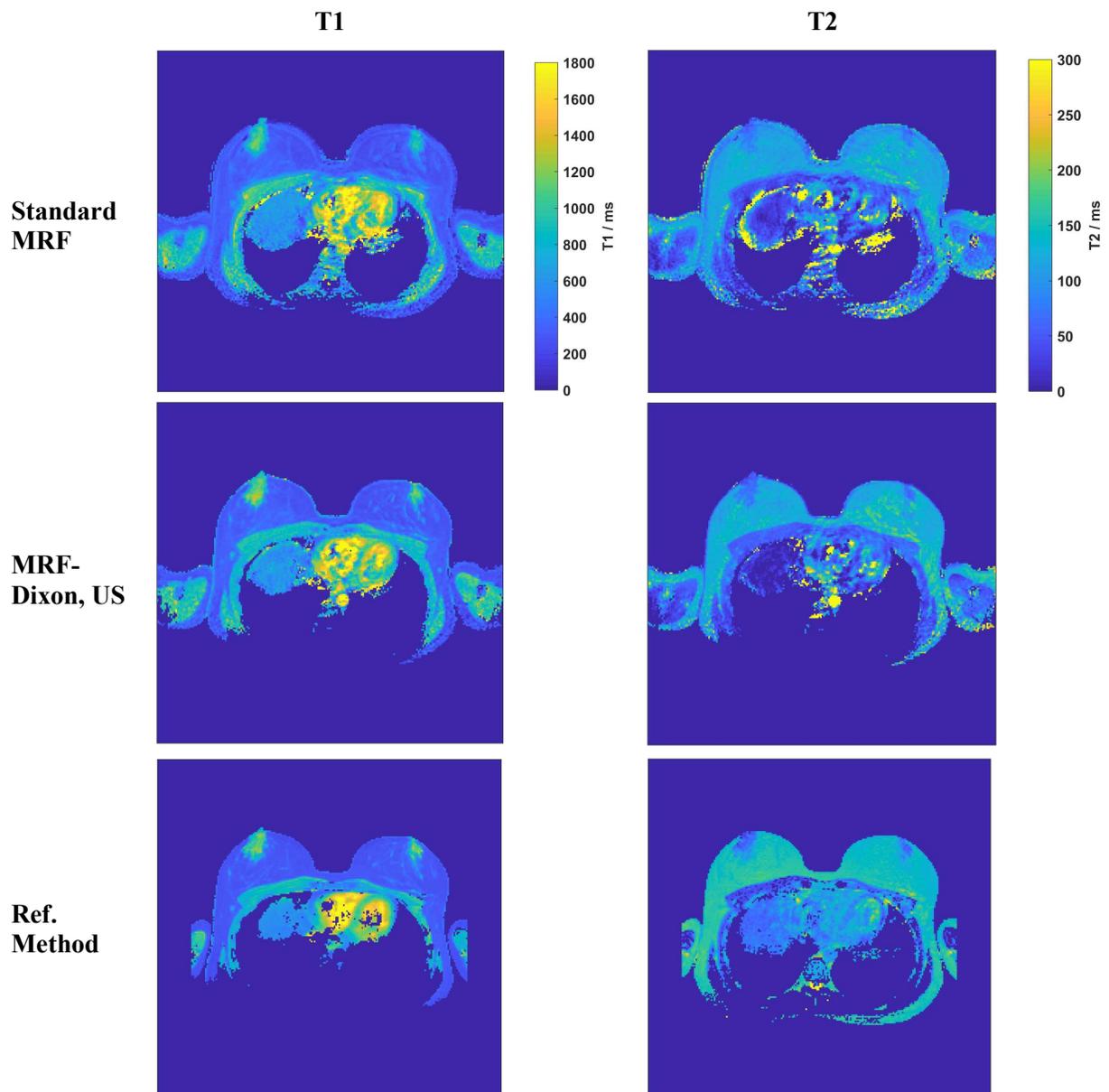

For volunteer 1, the fully sampled MRF-Dixon scan was not acquired.

(b) Volunteer 2

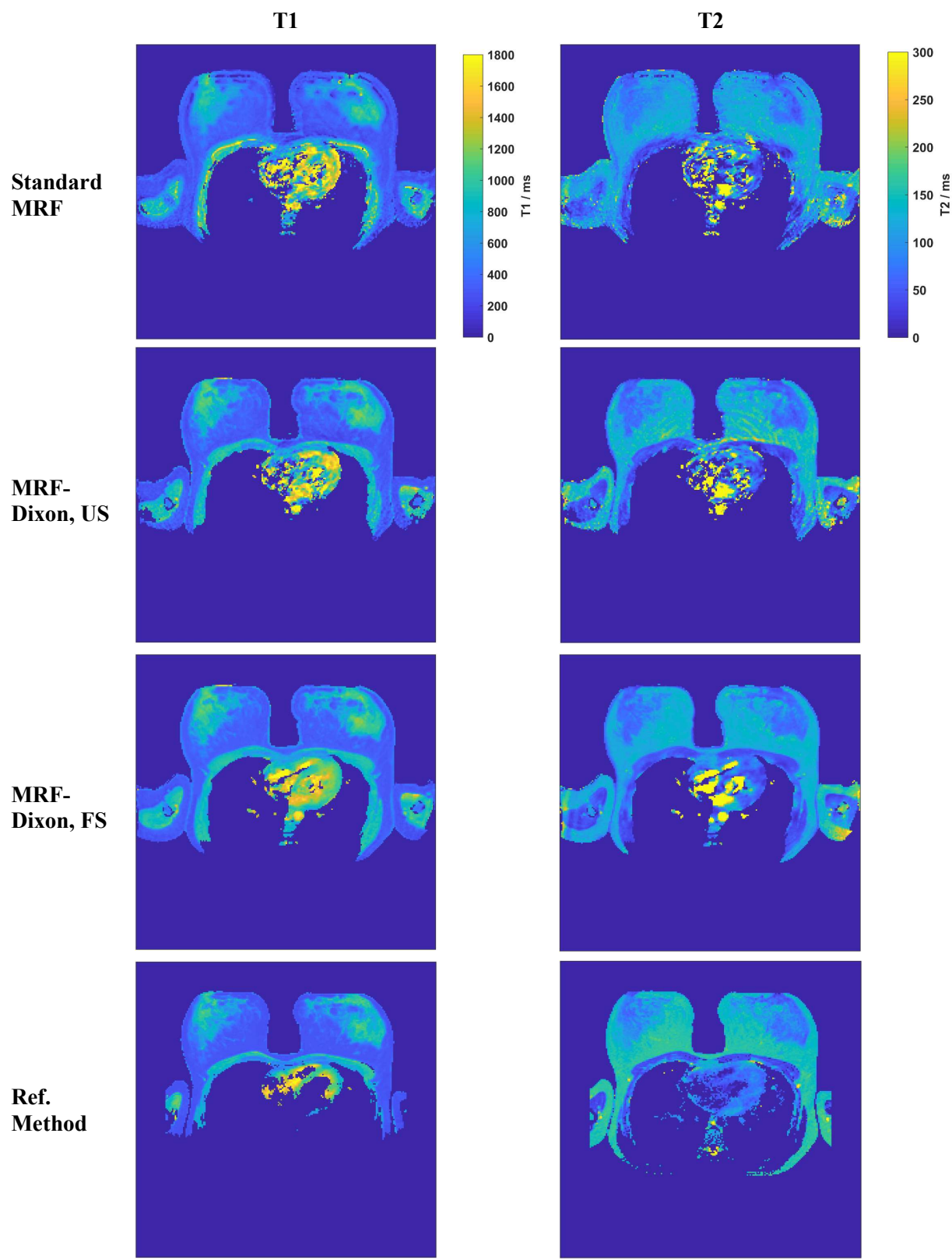

## (c) Volunteer 3

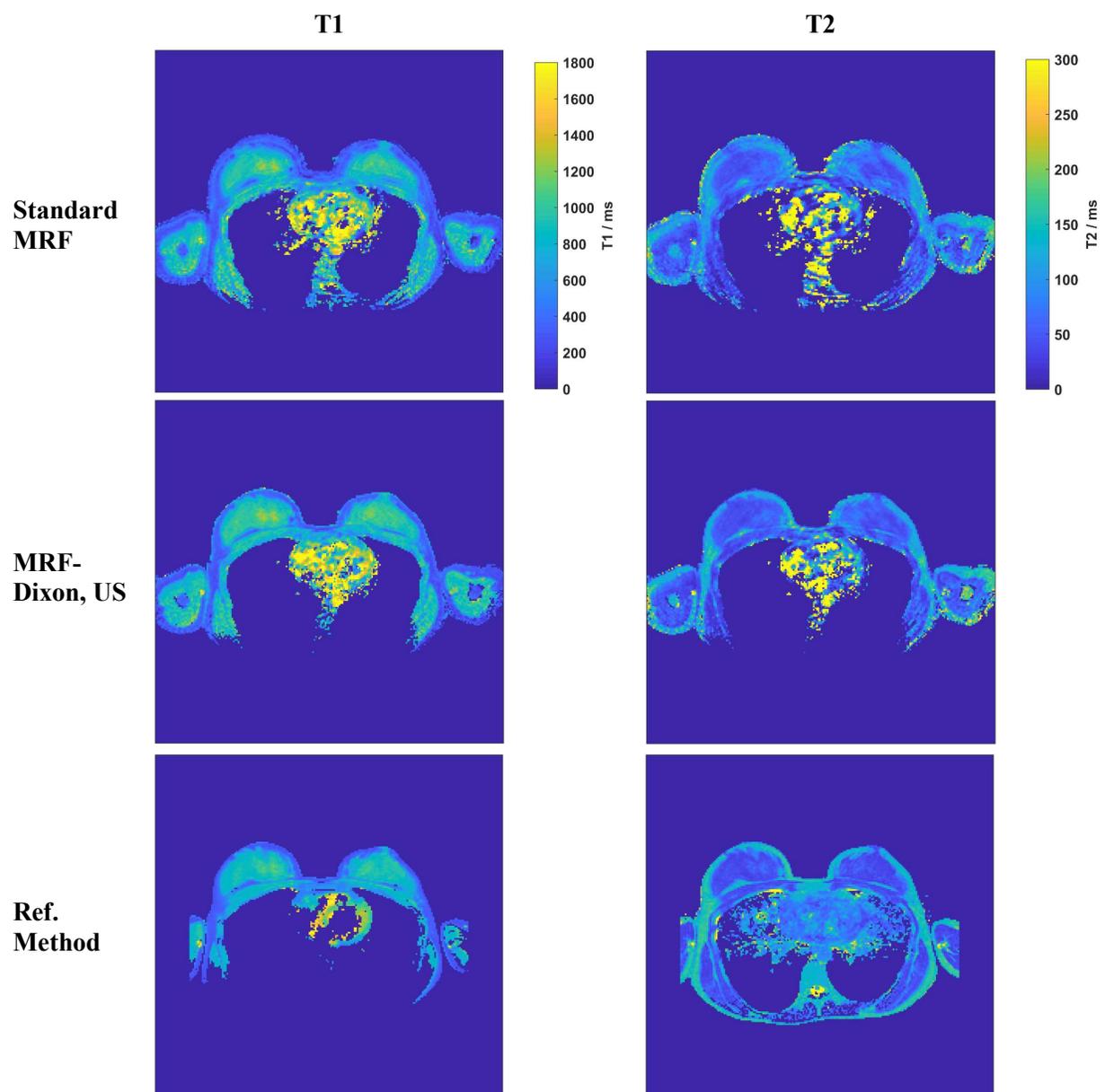

For volunteer 3, the fully sampled MRF-Dixon scan was not acquired.

(d) Volunteer 5

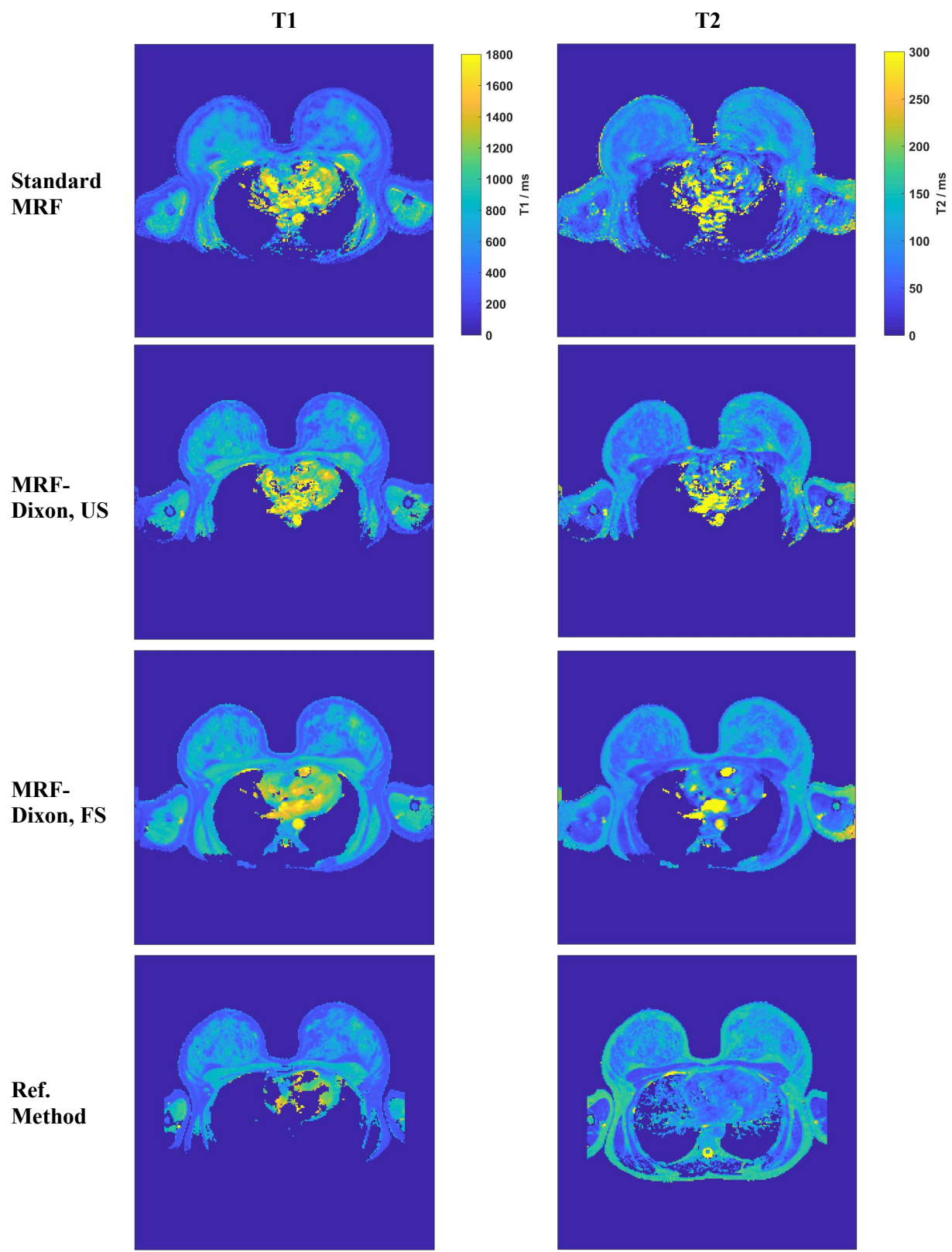

## (e) Volunteer 6

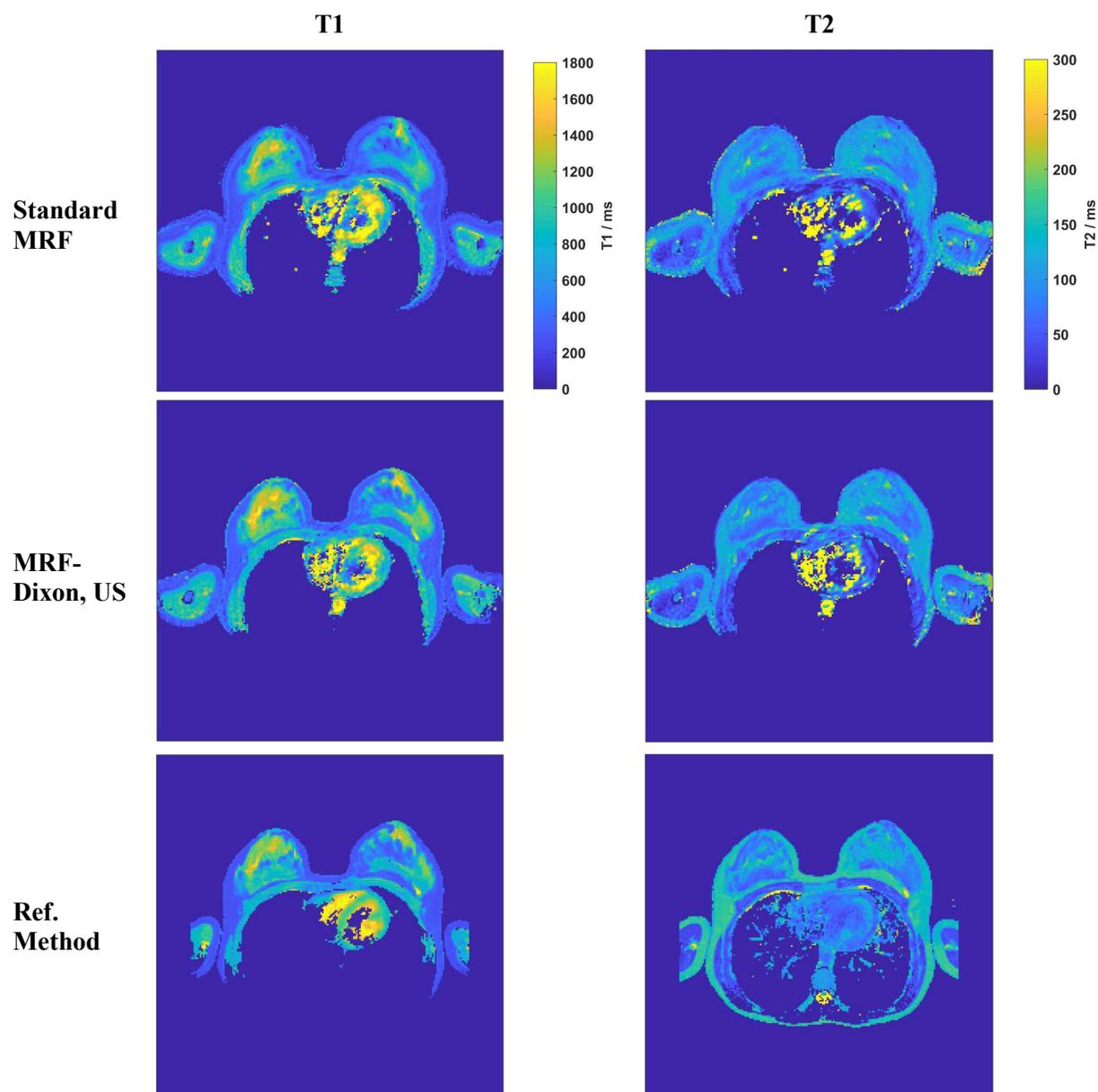

For volunteer 6, the fully sampled MRF-Dixon scan was not acquired.

## Acknowledgements

This project has received funding from the European Union's Horizon 2020 research and innovation programme under the Marie Skłodowska-Curie grant agreement No 642445 and under grant agreement No 667211.